\documentclass{aa}
\usepackage{natbib,graphicx}
\begin{document}

\title{Discovery of pulsations from the Be/X-ray binary
  \object{RX\,J0101.3$-$7211} in the SMC by XMM-Newton
\thanks{Based on observations with XMM-Newton, an ESA Science Mission  
with instruments and contributions directly funded by ESA Member
states and the USA (NASA).}}

\titlerunning{Be/X-ray binary system RX J0101.3$-$7211}

\author{M.\ Sasaki\inst{1} \and F.\ Haberl\inst{1}
  \and S.\ Keller\inst{2} \and W.\ Pietsch\inst{1}}

\authorrunning{Sasaki et al.}

\offprints{M. Sasaki, \email{manami@mpe.mpg.de}}

\institute{Max-Planck-Institut f\"ur extraterrestrische Physik,
  Giessenbachstra{\ss}e, Postfach 1312, 85741 Garching, Germany \and Lawerence
  Livermore National Laboratory, 7000 East Ave., Livermore, CA 94550-9234, USA}
     
\date{Received February 14th, 2001; accepted February 21th, 2001}

\abstract{
We report pulsations in the X-ray flux of \object{RX\,J0101.3$-$7211} in the
Small Magellanic Cloud (SMC) with a period of ($455\pm2$)~s in XMM-Newton 
EPIC-PN data. The X-ray spectrum can be described by a power-law with a photon 
index of $0.6\pm0.1$. Timing analysis of ROSAT PSPC and HRI archival data 
confirms the pulsations and indicates a period increase of $\sim5$~s since 
1993. \object{RX\,J0101.3$-$7211} varied in brightness during the ROSAT 
observations with timescales of years with a maximum unabsorbed flux of 
$6 \times 10^{-13}$~erg~cm$^{-2}$~s$^{-1}$ (0.1 -- 2.4\,keV). 
The flux during the XMM-Newton observation in the ROSAT band was lower than
during the faintest ROSAT detection. The unabsorbed luminosity derived from
the EPIC-PN spectrum is $2 \times 10^{35}$~erg~s$^{-1}$ (0.2 -- 10.0\,keV)
assuming a 
distance of 60~kpc. Optical spectra of the proposed counterpart taken at the 
2.3~m telescope of MSSSO in Australia in August 2000 show strong
H$\alpha$ emission and indicate a Be star. The X-ray and optical
data confirm \object{RX\,J0101.3$-$7211} as a Be/X-ray binary pulsar in
the SMC.      
}

\maketitle

\section{Introduction}

The investigation of X-ray binary populations in nearby galaxies allows to
study source samples at a similar distance. The
Magellanic Clouds which are heavily interacting with our Galaxy and
which show different metalicities are of particular interest in this respect.

Over the last decade observations of the Small Magellanic Cloud with
ASCA, Beppo-SAX, ROSAT and Rossi-XTE revealed a large number of X-ray pulsars
in this neighboring galaxy. The majority of these are accreting neutron stars
in a binary system with a Be star as mass donor. These so-called Be/X-ray
binaries form the major sub-class of high mass X-ray binary systems.
\citet{2000A&A...359..573H} list 47 Be/X-ray binaries (including candidates)
in the SMC from which 15 were known as pulsars at the time of publication of
their catalogue. From three more listed objects X-ray pulsations were reported
\citep{2001ApJ...548L..41C,2000PASJ...52L..73Y,2000PASJ...52L..37Y} and three
new pulsars were discovered recently
\citep{2000IAUC.7361....2Y,2000PASJ...52L..53Y,2000IAUC.7562}. 
Together with SMC\,X-1, a supergiant high mass X-ray binary pulsar, this
results in a total of 22 X-ray pulsars known in the SMC
\citep{2001astro.ph..2017M}. 

One of the Be/X-ray binary candidates proposed by \citet{2000A&A...359..573H}
is the X-ray source \object{RX\,J0101.3$-$7211} (source No 43 in Haberl \&
Sasaki, \citeyear{2000A&A...359..573H}) which may be related to the ASCA
source No 27 in \citet{2000ApJS..128..491Y}. The high variability of a factor
of $\sim$10 \citep{2000A&AS..147...75S} and the presence of an emission 
line object \citep{1993A&AS..102..451M} in the X-ray error circle makes this
interpretation most likely. 
In this letter we report the discovery of X-ray pulsations in the flux of
\object{RX\,J0101.3$-$7211} in XMM-Newton observations of the nearby 
SNR\,0102$-$72.3 in the SMC. Optical spectra taken at MSSSO 
confirm the pulsar as Be/X-ray binary.

\section{XMM-Newton Observations}

\object{RX\,J0101.3$-$7211} was detected serendipitously in XMM-Newton
\citep{2000SPIE.4012..731A} observations pointed at SNR\,0102$-$72.3 in April
2000 during the calibration phase of the satellite (Table\,\ref{obslist}). 
It was located in the EPIC-PN \citep{2001A&A...365L..18S} image, but outside 
the field of view of the EPIC-MOS cameras \citep{2001A&A...365L..27T}.
The two EPIC-PN observations were carried out in full frame mode with thin and
medium filter inserted and the data processed with the XMM-SAS software. 
A comparison of coordinates derived from EPIC-PN data and the more accurate
ROSAT HRI positions for sources which were identified with known objects
resulted in a discrepancy of $\sim10$\arcsec\,. Therefore the coordinates
of the EPIC-PN images were corrected using the HRI coordinates as reference. 
The remaining $\sim6$\arcsec\ statistical uncertainty in the EPIC-PN position 
of \object{RX\,J0101.3$-$7211} makes the identification with the ROSAT source 
most likely. The events from the source 
were selected from both observations resulting in effective exposure times of
$\sim$18\,ks each. The background for the spectral analysis was estimated from
regions at 40\arcsec\ to 90\arcsec\ distance from the source.  

\subsection{X-ray spectrum}

X-ray spectra of \object{RX\,J0101.3$-$7211} were extracted from thin and 
medium filter data separately and both vignetting corrected. The two spectra 
were fitted simultaneously with a power-law model using XSPEC with the Nov.\ 
2000 version of the response matrix. We fixed the foreground absorption at the
Galactic $N_{\rm H\,gal}$ of $5.8 \times 10^{20}\,\mathrm{cm}^{-2}$
\citep{1990ARA&A..28..215D}, and fitted an additional absorption column
density with fixed elemental abundances of 0.2 solar to account for the
interstellar gas in the SMC \citep{1992ApJ...384..508R}. 
The best fit was achieved with a photon index $\Gamma = 0.6\pm0.1$
and SMC absorption of $N_{\rm H\,int} = 1.3^{+2.8}_{-1.3} \times
10^{21}$~cm$^{-2}$. 
The unabsorbed flux calculated from the thin filter data is $4.2 \times
10^{-13}$~erg~cm$^{-2}$~s$^{-1}$ (0.2 -- 10.0\,keV), with a luminosity of 
$L_{\rm X} = 1.8 \times 10^{35}$~erg~s$^{-1}$ for an assumed distance of
60\,kpc. 

\begin{figure}
\centerline{\resizebox{7.5cm}{!}{\includegraphics[angle=270,width=8cm,clip,bb=81 38 570 705]{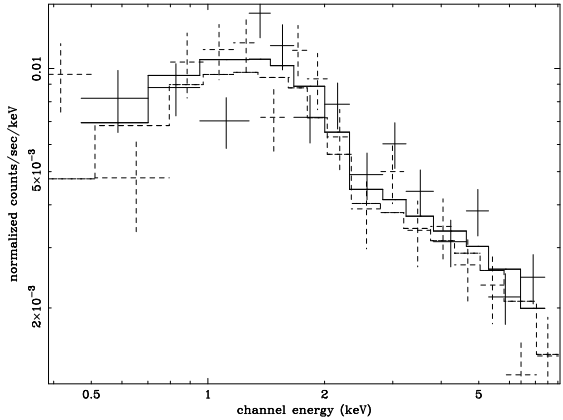}}}
\caption[]{\label{spectrum} XMM-Newton EPIC-PN spectrum and the fitted
  power-law model. Solid line is used for thin filter data and fitted model,
  dashed line for medium filter data and model}
\end{figure}

\subsection{Variability}

\begin{figure}
\centerline{\resizebox{8cm}{!}{\includegraphics[angle=270,clip,bb=60 30 525 700]{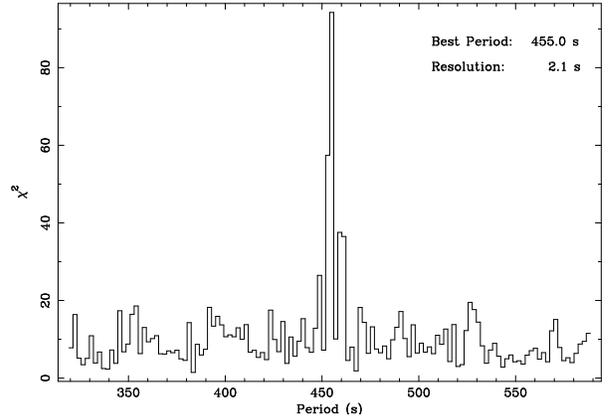}}}
\caption[]{\label{period} $\chi^2$-test periodogram for
  \object{RX\,J0101.3$-$7211} 
  from EPIC-PN thin and medium filter observations}
\end{figure}

\begin{figure}
\centerline{\resizebox{8cm}{!}{\includegraphics[angle=270,clip,bb=70 30 540 700]{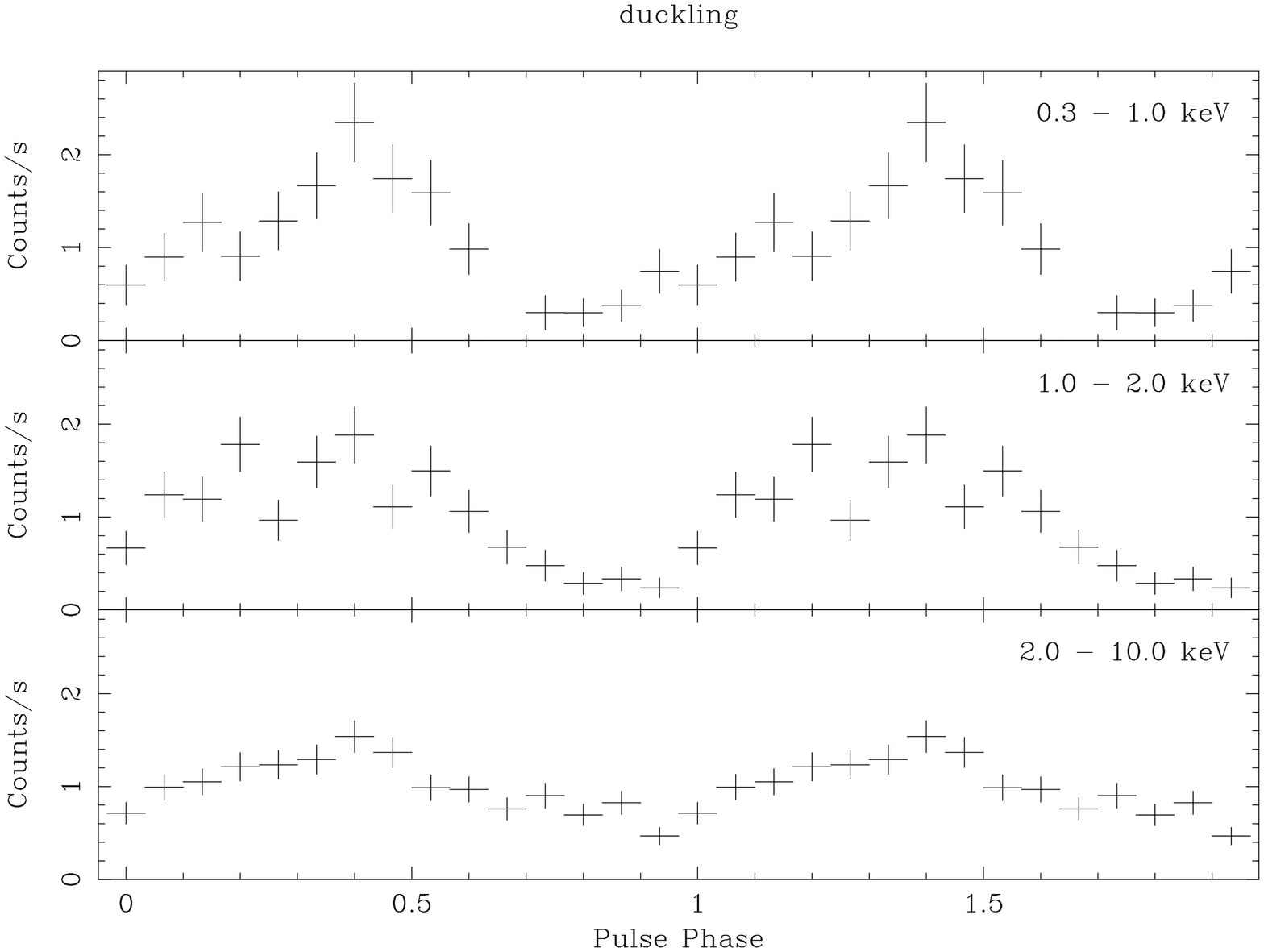}}}
\caption[]{\label{folded} Folded lightcurves of \object{RX\,J0101.3$-$7211}
in three energy bands derived from EPIC-PN thin and medium filter 
observations with binning time of 30.33~s. Phase 0 is 16 Apr.\ 2000
19:56:48.9 (UTC)} 
\end{figure}

The XMM-Newton EPIC-PN observations with a time resolution of 73.3\,ms
\citep{2001A&A...365L..18S} allow to search
for pulsations in the X-ray flux of \object{RX\,J0101.3$-$7211} which was
known as Be/X-ray binary candidate \citep{2000A&A...359..573H}. 
Source events from both thin and medium filter observations in the
energy band of 0.2 -- 10.0~keV were analyzed with XRONOS FTOOLS.
Based on power spectra and following detailed period search in a time interval 
of 330 -- 590~s with a resolution of 2.1~s revealed pulsations with a period of
$P = (455\pm2)$\,s (see Fig.\,\ref{period}). The data was separated in three
energy bands in order to analyze the energy dependence of the flux variation: 
0.3 -- 1.0~keV (soft), 1.0 -- 2.0~keV (medium), and 2.0 -- 10.0~keV (hard). 
As can be seen in the folded lightcurves of the different bands in
Fig.\,\ref{folded} including thin and medium observations, the
pulsations in the medium and hard bands are similar, whereas in the soft band,
the minimum comes earlier than in the harder bands. The maximum is
reached almost at the same time in all three bands. Thus the increase in the
soft band is significantly flatter and the decrease steeper than in the medium
and the hard band. 

\begin{table*}
\caption[]{\label{obslist} ROSAT and XMM-Newton observations of
  \object{RX\,J0101.3$-$7211}}  
\begin{tabular}{lllcccccl}
\hline
\noalign{\smallskip}
Telescope & Detector & Obs.\ ID & Start Time & End Time & \multicolumn{2}{c}{Pointing Direction} & Exposure & Filter \\
 & & & & & RA & Dec & [ks] & \\
\hline\hline
\noalign{\smallskip}
ROSAT & PSPC & 400300p & 29 Mar.\ 1993 & 30 Mar.\ 1993 & 00 58 33.6 & $-$71
36 00 & 5.2 & \\
& & 500142p & 12 May 1993& 13 May 1993& 01 04 02.4 & $-$72 01 48 & 4.9 & \\
& & 400300p-1 & 01 Oct.\ 1993& 09 Oct.\ 1993& 00 58 33.6 & $-$71 36 00 &
 7.2 & \\
ROSAT & HRI & 500137h & 17 Apr.\ 1993 & 21 Apr.\ 1993 & 01 03 16.8 & $-$72
09 36 & 14.1 & \\
& & 900445h & 19 Apr.\ 1994 & 09 Jun.\ 1994 & 00 59 28.8 & $-$72 09 36 &
14.9 & \\  
& & 900445h-1 & 13 Apr.\ 1995 & 12 May 1995 & 00 59 28.8 &
$-$72 09 36 & 34.6 & \\
& & 500418h & 31 May 1995 & 02 Jun.\ 1995 & 00 59 26.4 & $-$72 10 12 & 2.0
& \\ 
& & 500418h-1 & 15 Oct.\ 1995& 30 Oct.\ 1995& 00 59 26.4 & $-$72 10 12 &
 8.3 & \\
& & 500418h-2 & 11 May 1997& 10 Jun.\ 1997& 00 59 26.4 & $-$72 10 12 &
 7.5 & \\
& & 500418h-3 & 27 Mar.\ 1998& 01 Apr.\ 1998& 00 59 26.4 & $-$72 10 12 &
 11.0 & \\
XMM-Newton & EPIC-PN & 0123110201 & 16 Apr.\ 2000 & 17 Apr.\ 2000 & 01
03 50.0 & $-$72 01 55 & 18.8 & thin\\
& & 0123110301 & 17 Apr.\ 2000 & 17 Apr.\ 2000 & 01 03 50.0 & $-$72 01
 55 & 18.3 & medium\\
\hline
\end{tabular}
\end{table*}

\section{Long-term X-ray variability}

\subsection{ROSAT observations}

Searching the ROSAT \citep{1982AdSpR...2..241T} archive, six pointed
observations with PSPC \citep{1987SPIE..733..519P} and HRI
\citep{1996rosat.hri...D} as focal instruments were found in which  
\object{RX\,J0101.3$-$7211} was detected. Further four pointings
yield upper limits for the source flux (Table\,\ref{obslist}). 
The count rates and upper limits
were determined using the source detection routine of EXSAS 
\citep{1994exsas.....Z}. 
In order to compare the count rates of the PSPC and HRI detectors, the PSPC
count rate was converted into HRI count rate by dividing by a factor of 3.0,
a typical value for sources with a hard X-ray spectrum like that of
X-ray binaries \citep{2000A&AS..143..391S}.
 
\subsection{Flux variations}

\begin{figure}
\centerline{\resizebox{8cm}{!}{\includegraphics[angle=270,clip,bb=40 75 570 780]{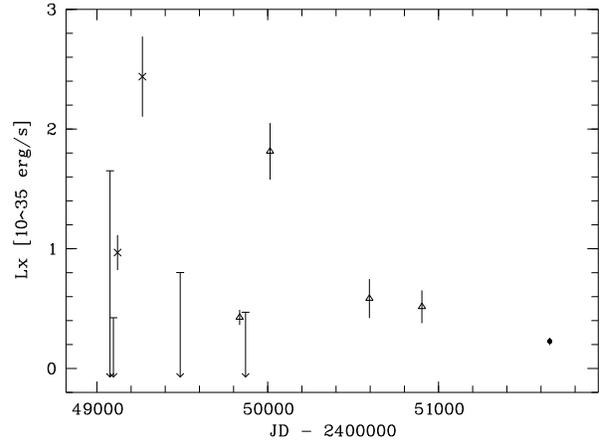}}}
  \caption[]{\label{longterm} X-ray lightcurve of \object{RX\,J0101.3$-$7211}
    derived  
    from ROSAT PSPC (marked with {\sf x}), HRI (triangles)  and XMM-Newton
    EPIC-PN (last data point) data. Upper limits
    determined from ROSAT observations are plotted with arrows}
\end{figure}

Variability of \object{RX\,J0101.3$-$7211} was reported by
\citet{2000A&AS..147...75S} based on ROSAT PSPC and HRI observations. 
Fig.\,\ref{longterm} shows the lightcurve including both ROSAT PSPC and HRI
observations, as well as the new EPIC-PN data.
The luminosity was calculated assuming the same spectrum for all observations
as determined from the EPIC-PN data and extrapolated to the ROSAT band, and
a distance of 60\,kpc.  
Long term variation can been seen with a maximum unabsorbed flux of $5.7 \times
10^{-13}$~erg~cm$^{-2}$~s$^{-1}$ (0.1 -- 2.4\,keV). 
The unabsorbed flux during the XMM-Newton observation in the ROSAT band
was $5.9 \times 10^{-14}$~erg~cm$^{-2}$~s$^{-1}$ corresponding to a luminosity
of $L_{\rm X} = 2.5 \times 10^{34}$~erg~s$^{-1}$, about a factor of two lower
than during the faintest ROSAT detection.

\subsection{Pulse period history}

Archival ROSAT PSPC and HRI data were analyzed to look for pulsations in
earlier observations. Knowing the period from XMM-Newton observations 
allowed us to restrict the search around the expected value.
Two observations (400300p-1, when the source was brightest and
900445h-1, the longest ROSAT observation) provided sufficient source
counts, that a Z$^2$ test yielded pulse periods of $P = (450.2\pm1.5)$\,s and
(452$\pm$3)\,s, respectively. The large errors are caused by possible aliases
due to the interrupted ROSAT observations. The derived pulse periods indicate
a period increase of $\sim$5~s since Oct.\ 1993. This corresponds to a spin
down rate of $\dot{P} = (2.3\pm2.0) \times 10^{-8}$~s~s$^{-1}$ (90\%
confidence).  

\section{Optical counterpart}

\begin{figure*}
\resizebox{12cm}{!}{\includegraphics[angle=270,bb=40 35 570 800,clip]{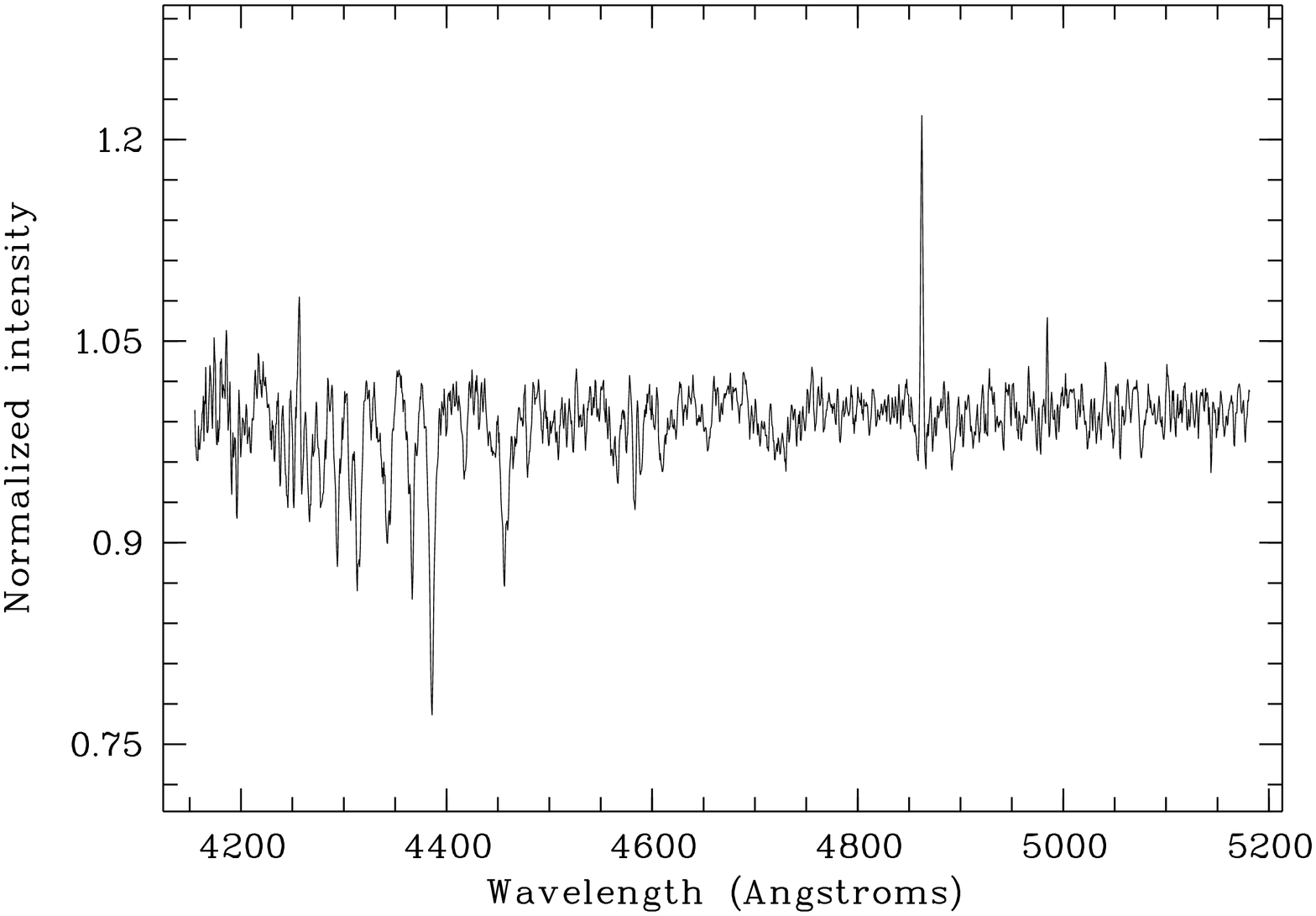} 
\includegraphics[angle=270,bb=40 110 570 780,clip]{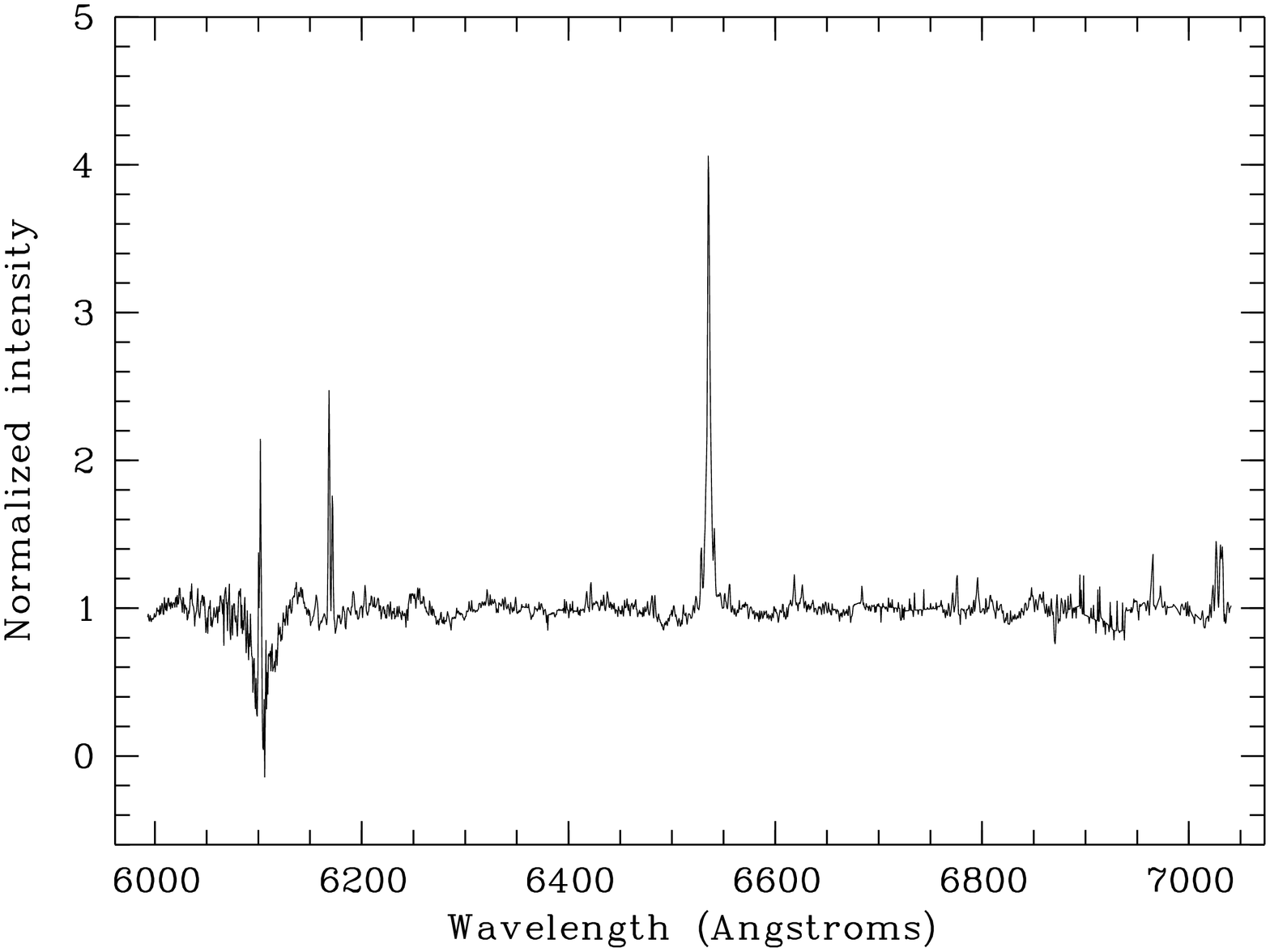}
}
\hfill
\parbox[t]{55mm}{
  \caption[]{\label{optspec} Spectrum of the optical counterpart
taken at the 2.3\,m telescope of MSSSO in Australia in August 2000.}
}
\end{figure*}

An emission line object from the catalogue of \citet{1993A&AS..102..451M} 
was suggested as optical counterpart for \object{RX\,J0101.3$-$7211} 
by \citet{2000A&A...359..573H}. A medium resolution (0.6\,\AA~px$^{-1}$)
spectrum of the star was obtained on 20 August 2000 
using the Double Beam Spectrograph on the MSSSO 2.3~m in Australia. The
spectra consist of two simultaneously recorded non-overlapping segments: blue
(4150 - 5200)\,\AA\ and red (6200 - 6800)\,\AA, as shown in
Fig.\,\ref{optspec}. The segments were wavelength
calibrated using a CuAr arc. The red segment shows prominent H$\alpha$
emission (equivalent width 11\AA). The blue segment shows weak H$\beta$
emission and the absence of H$\gamma$ absorption due to emission infill of the
line. Lines of He\,I are seen most prominently at 4471\,\AA\ and 4388\,\AA.

\section{Discussion}\label{discussion}

\object{RX\,J0101.3$-$7211} was detected in ROSAT observations and 
proposed as Be/X-ray binary candidate. In the error circle of the ROSAT HRI
data with a radius of 3\farcs6 \citep[source No
95 in][]{2000A&AS..147...75S} an emission line object was suggested by
\citet{2000A&A...359..573H} as counterpart. 
Optical follow-up observations presented here show a spectrum
with strong H$\alpha$ emission confirming the Be star nature of the proposed 
counterpart.

The X-ray spectrum obtained from the XMM-Newton EPIC-PN data is well described
by a power-law model ($\Gamma = 0.6\pm0.1$) with an absorption of
$N_{\rm H\,int} = 1.3^{+2.8}_{-1.3} \times 10^{21}$~cm$^{-2}$ besides to the
Galactic foreground absorption. The luminosity
in the energy range of 0.2 -- 10.0\,keV was $1.8 \times
10^{35}$~erg~s$^{-1}$ indicating a faint phase. X-ray luminosities determined
for Galactic and Magellanic Cloud X-ray binaries normally range from $L_{\rm
  X} \approx 10^{34}$~erg~s$^{-1}$ to $10^{38}$~erg~s$^{-1}$
\citep[and references therein]{1994SSRv...69..255A, 2000A&A...359..573H}. The
spectrum of the nearby ASCA source which probably coincides with
\object{RX\,J0101.3$-$7211} was also described with a power-law model with
$\Gamma = 0.6$ and a total $N_{\rm H}$ of $1 \times 10^{21}$~cm$^{-2}$
\citep{2000ApJS..128..491Y}, consistent with the EPIC-PN spectrum. The ASCA
source was brighter with a luminosity of $7.3 \times 10^{35}$~erg~s$^{-1}$
(0.7 -- 10.0\,keV) which would indicate a bright state, but no pulsations were
found \citep{2000ApJS..128..491Y}. 

We discovered pulsations in the X-ray flux of \object{RX\,J0101.3$-$7211} with
a period of $P = (455\pm2)$\,s in EPIC-PN data, indicating the spin
period of a neutron star. Therefore \object{RX\,J0101.3$-$7211} is one of the
X-ray binaries with longer spin periods detected in the SMC (see Haberl \&
Sasaki, \citeyear{2000A&A...359..573H} and references therein). The longest
spin period of a SMC Be/X-ray binary 
measured so far is ($755.5\pm0.6$)~s for AX\,J0049.5-7323
\citep{2000PASJ...52L..73Y}. The analysis of archival ROSAT data suggest a
spin period increase of $\dot{P} = (2.3\pm2.0) \times
10^{-8}$~s~s$^{-1}$ over 
the last seven years. Similar values for spin up and spin down have been
reported for the compact object in various Be/X-ray binary systems 
\citep{1994SSRv...69..255A}. The X-ray results combined with the optical
spectrum assure that \object{RX\,J0101.3$-$7211} is an X-ray binary in the SMC
with a neutron star orbiting a Be star companion.

\begin{acknowledgements}
The XMM-Newton EPIC-PN data were kindly made available by the RGS team (A.\
Rasmussen et al.). 
The XMM-Newton project is supported by the Bundesministerium f\"ur
Bildung und Forschung / Deutsches Zentrum f\"ur Luft- und Raumfahrt
(BMBF/DLR), the Max-Planck Society and the Heidenhain-Stiftung.
\end{acknowledgements}
   
\bibliography{/home/manami/tex/bibtex/my,/home/manami/tex/bibtex/xb,/home/manami/tex/bibtex/lmchricat2000,/home/manami/tex/bibtex/smchricat2000}

\end{document}